\magnification1200

\vskip 2cm
\centerline {{\bf The discovery of supergravity and its early development } \footnote {*} {To appear in the 40th Anniversary Special Issue of International Journal of Modern Physics A and Modern Physics Letters A. }}

\vskip 1cm
\centerline{\bf  Peter West }
 \vskip 1.2cm
	
\vskip 0.5cm
\centerline{{\it Mathematical Institute, University of Oxford,}}
\centerline{{\it Woodstock Road, Oxford, OX2 6GG, UK}}
\centerline{}
\centerline{{\it Department of Mathematics, King's College, London}}
\centerline{{\it The Strand, London WC2R 2LS, UK}}

\vskip 1.7cm

\leftline{\sl Abstract}
I give an account of the discovery of supergravity and the first six years of its  development which lead to the most general coupling of supergravity to matter.  I  restrict the discussion to theories with only one supersymmetry that are in four dimensions. As well as describing the contents of the papers I will  also  discuss the ideas and techniques they used. I begin by giving an account of the  discovery of  supersymmetry. 
\noindent
\vskip 6cm
email: peter.west540@gmail.com
\vfill

\eject

\medskip
{\bf 0. Preface}
\medskip
Aware of  the approach of the fiftieth anniversary of the discovery of supergravity,  and at the urging of others,   I will give  a first hand account of the  discovery of supergravity and its very early development. I will restrict my attention to theories in four dimensions that have simple supersymmetry, that is,  $N=1$ supersymmetry which has  only four supersymmetry generators. I will end with  the construction of the most general such theories which was achieved in 1982.  Apart from a few later pedagogical references I will only reference papers published during this period. 
\par
The discovery of physics is a human endeavour. The path ahead is not always clear and   papers which  make major break throughs can go unread for years. This is well illustrated by the development of supersymmetry and supergravity. As well as describing  the contents of the papers I will also explained the ideas and methods the authors used as  these are also likely to  be of use in the future. The history of science  can sometimes be complicated and  it not always easy to get it right. Although I have made a detailed study of the papers during this period  I would welcome any corrections. 
\par
 During this quest,  and partly out of curiosity,  I also began reading the early papers on supersymmetry. I was surprised to find that what I thought they contained and what they did contain was not the same. As such it became apparent to me that the contents of some of the key papers are not that well known. As a result,  I will first  discuss the discovery of supersymmetry.

\medskip
{\bf 1. The discovery of supersymmetry}
\medskip
We will begin by discussing the discovery of supersymmetry. which was discovered independently in two very different ways. In 1971  Golfand and Liktman, published a remarkable paper [1]. It contained on its  first page a new type of symmetry algebra that had anti-commutators as well as the usual commutators. This algebra contained the Poincare algebra as well as a spinor generator  that obeyed the anti-commutators instead of the previously  used commutators. The authors stated that  " In order not to violate ... the connection between the spin and statistics, we shall  take  anti-commutators...." for the new generators. The algebra Golfand and Liktman wrote down was that of $N=1$ supersymmetry. The mathematical framework to incorporate anti-commuting objects and anti-commutation relations  in quantum field theory was worked out by Berezin in mid-to late 1960's.
\par
Although only briefly alluded to in their paper, there had been a quest to find a symmetry that incorporated in a non-trivial way the Poincare group with  an internal symmetry group. The hope had been that if such a symmetry  existed it could only contain certain internal symmetry groups and this would narrow down the possible internal symmetries that could  occurr in nature. These hopes were apparently dashed when Coleman and Mandula  [2]  showed that if a theory had such a symmetry  then the  particle scattering would be trivial. The hidden assumption was that the generators of the symmetry obeyed  a Lie algebra which always had commutators. 
\par
On the second page of their paper they discussed two representations of their new algebra. One had two spin zero  and one spin one half particles all of which were massive. The other had one spin one half particle, one spin zero particle and one spin one particle all of which were also  massive. From the view point of later developments these were the $N=1$ chiral multiplet and the $N=1$ massive spin one multiplet. The transformations of these particles under supersymmetry was not given but the  contribution to the supercharge for each representation was presented and from these it would have been easy to compute the transformations. 
\par
On page three of the Golfand and Liktman  paper there appeared  a four dimensional model which was invariant under supersymmetry. It contained both of the above representations of supersymmetry. It was an interacting theory with one  constant determining  the interaction between the two representations.  Their paper  stated that their model broke parity,  but this was   just a misunderstanding and it did not affect the results in the paper. Yuri Golfand was the only person chosen to be sacked from the  Lebedev Physics Institute in 1973.   
\par
The paper of Golfand and Liktman attracted little attention and essentially no attention in the west. At that time travel to and from  Moscow was very limited and when it did occur it was carefully documented. This required substantial administration from those Russians connected with the travel.  Also the process to publish papers in Soviet Union was not straightforward and could be  lengthy. This had the advantage that those in Russia could develop their own ideas free from any distraction and supersymmetry was one of a number of such ideas that were to become crucial for the development of theoretical physics. 
\par
A bit later, in 1972, Volkov and Akulov  also independently proposed a supersymmetric model [3]. Their two page paper was entitled "Is the neutrino a Goldstone particle". They introduced a spinor field $\lambda_\alpha$  which depended on the usual coordinate of spacetime $x^\mu$. They  wrote down a transformation with an anti-commuting spinor parameter  $\epsilon_\alpha$ mixed the spinor field  $\lambda_\alpha$ and the coordinates of spacetime, namely 
$$
 x^{\mu\prime}= x^\mu -\bar \epsilon \gamma^\mu \lambda \ ,\  \lambda_\alpha ^\prime=  \lambda_\alpha+ \epsilon_\alpha 
\eqno(1.1)$$
They stated that the closure of these transformations and those of the Poincare group was an algebra  with anti-commuting parameters. They did not, however, write down this algebra but  it was the $N=1$ supersymmetry algebra. It is remarkable how such a simple equation foreshadowed so much of the future  development  of supersymmetry. 
In this paper Volkov and Akulov took  $\lambda_\alpha$ to be a field rather than a anti-commuting  coordinate,  but they did take it to be a coordinate  in a later paper [4] on the same subject. This was the superspace,  independently discovered and  extensively developed by Salam and Strathdee [5]. 
\par
They then constructed a theory,  in four dimensions,  that was invariant under  the super Poincare group,  but this was spontaneous broken  to preserve only the Poincare subgroup.     The breaking of the supersymmetry  lead to a Goldstone fermion  which was a massless spin one half particle that  they proposed could be the neutrino. From a modern perspective they had constructed the  non-linear realisation of the super Poincare group in which supersymmetry was broken. In appendix A1 we give an account of this construction. This  construction must have been  natural for  Volkov who was one of the pioneers of non-linear realisations, especially those that involved the generators associated with spacetime. 
\par
Supersymmetry was also  hidden in the superstring found by Rammond [6] and Neveu and Schwarz [7].  The physical states  of the bosonic string  were defined by certain operators acting on the vacuum of  the theory. These operators were constructed out of oscillators which in turn arose, upon quantisation,  from the fields, $x^\mu (\tau , \sigma)$,  in the two dimensional world sheet theory that described the string. These fields described the position of the string and classically obeyed the Nambu action, but from the two dimensional viewpoint they were scalar fields. For the bosonic string these  operators generated the Virasoro  algebra. 
\par
The Virasoro  generators were also present in the superstring, but  the superstring contained in addition  some other operators that obeyed anti-commutators. All these generators were required to define the physical states of the superstring. 
In 1971 Gervais and Sakita [8] realised that the two dimensional world sheet description of the superstring was a two dimensional field theory that contained, from the two dimensional viewpoint,  spin zero particles  ($x^\mu (\tau , \sigma)$),  but also spin one half fermions  ($\psi^\mu (\tau , \sigma)$). These latter anti-commuting  fields were required  to give, upon quantisation, the fermionic oscillators needed to construct the additional fermionic generators required to define the physical states of the superstring. 
\par
Gervais and Sakita wondered if there was a symmetry between the spin zero and spin one half fields of the two dimensional  world sheet dimensional theory. Suppressing the spacetime index $\mu$, they found that the free two dimensional action 
$$
{1\over 2}\int d^2 x \{ -\partial_\mu A\partial^\mu A -i\bar \chi \gamma^\mu \partial_
\mu \chi + F^2\}
 \eqno(1.2)$$
was invariant under the transformations 
$$
\delta A= i\bar \epsilon \chi ,\ \delta \chi= (\gamma^\mu \partial_\mu A+ F)\epsilon ,\ \delta F= i \bar \epsilon \gamma^\mu \partial_\mu \chi
\eqno(1.3)$$
The auxiliary field $F$ was later introduced by Wess and Zumino and was presented in a lecture by Zumino  [9]. 
\par
 It was this model that Wess and Zumino generalised to four dimensions to construct what became  known as the Wess-Zumino model  [10] . This interacting field theory had the well known action
  $$
 A=\int d^4x\left\{-{1\over2}(\partial_\mu
A)^2-{1\over2}(\partial_\mu
B)^2-{1\over2}\bar\chi/\!\!\!\partial\chi+{1\over2}F^2+{1\over2}G^2\right\}
$$
$$
-m(AF+GB-{1\over 2}\bar{\chi}\chi )-\lambda\{(A^2-B^2)F+ 2GAB-\bar{\chi}(A-i\gamma_5B)\chi\}
\eqno(1.4)$$
  which was invariant under the supersymmetry transformations 
$$
\delta A=\bar\varepsilon\chi\ \ \delta B=i\bar\varepsilon\gamma_5\chi , \ \ 
\delta\chi=[F+i\gamma_5G+/\!\!\!\partial(A+i\gamma_5B)]\varepsilon
$$
$$
\delta F=\bar\varepsilon /\!\!\!\partial \chi , \  \ \delta G= i\bar\varepsilon\gamma_5 /\!\!\!\partial \chi
\eqno(1.5)$$
 which formed a closed algebra, namely, the $N=1$ supersymmetry algebra.  
 \par
 Not only did Wess and Zumino generalise the supersymmetric model to four dimensions  they also showed that  supersymmetry allowed  the theory to have  interactions and these  were  largely determined by this symmetry. They also noted that their model had fewer divergences than one might expect from such a generic quantum field theory. This made it clear, to those inclined to listen,  that supersymmetry was a powerful symmetry which could be of  relevance to the world in which we live. It helped that the  paper had the usual clarity of thought and elegance generic to  Wess-Zumino papers.  
 \par
 Their  paper did spark interest  in supersymmetry in a small number of largely European institutions. It was game on, the construction of the possible supersymmetric theories   in four dimensions began. These were models in which the supersymmetry transformations were rigid transformations, that is, their parameters were constants and so did not depend on spacetime. An account of these early developments of supersymmetry can be found in reference [B.1]. 
\medskip
{\bf 2 Supergravity}
\medskip
In this section we will explain how supergravity theory was discovered. As we mentioned we will restrict our attention to simple supergravity, that is, $N=1$ supergravity. 
\medskip
{\bf 2.1 The discovery} 
\medskip
The outstanding problem in 1974 was to find a theory that contained gravity and was supersymmetric.  It was clear that in addition to the spin two graviton one required an adjacent spin and a spin three-halves was the preferred candidate. The kinetic term for a free spin three-halves particle had been constructed by Rarita and Schwinger in 1941,  Buchdahl  coupled it to gravity in  1958 and even showed that for the coupling to be consistent gravity must obey Einstein's equation.  Velo and Zwanziger showed in 1969 that when the spin three-halves field was   coupled to the electromagnetic field it did not  propagate causally.  It was also apparent that 
the supersymmetry transformations must be local transformations as the commutator of two such  transformations must lead to a translation which,  in a gravity theory,  must  be a local transformation. 
\par
Superspace was independently  discovered by Salam and Strathdee [5] who introduced superfields together with  the geometry of superspace.  This gave a way to describe supersymmetric theories in such a way that their supersymmetry was manifest. Superspace is the coset of the Super Poincare group modulo the Lorentz  group and it has the coordinates $x^\mu$ and $\theta_\alpha$ which arise in the group element $g=e^{x^\mu P_\mu+ \theta_\alpha Q^\alpha}$. The Lorentz part of the super Poincare group element is not present as it can be removed by a Lorentz transformation  associated with the coset. 
\par
It was natural to apply  the ideas of Riemanian geometry in superspace in an attempt to find the sought after supergravity theory. The first paper in this direction was by Arnowitt, Nath and Zumino [11]. It closely mirrored the construction of general relativity and so it constructed Christoffel symbols and their Riemann curvature in superspace. 
\par
In 1975 Akulov, Volkov, and Soroka [12] also tried to construct supergravity in superspace in two papers,  but they introduced  curvatures and torsions. In an important step in the good direction they realised that the tangent space group in superspace was just the Lorentz group and so one could independently place constraints on components of the curvature and torsion without setting them to completely vanish. They also  explained how the superspace of Salam and Strathdee, used in the absence of supergravity,  had non-zero torsion and so this must be the case for supergravity. The formulation of supergravity in superspace had to wait until it had been formulated,  with its auxiliary fields,  in ordinary space
\par
Another attempt to construct supergravity was given  by Volkov  and Soroka [13,14] in 1973. These papers extended the approach of Volkov and Akulov,  discussed above,  to a theory which possessed local supersymmetry. They constructed the low energy effective action of a theory of local supersymmetry  that was spontaneously  broken to preserve Lorentz symmetry. They included in their considerations a local internal symmetry,  but this played no real role and so we will omit any mention of it here. More precisely, they constructed a non-linear realisation that introduced Goldstone fields for translations and supersymmetry transformations, but unlike before,  these were subject to local rather than rigid super Poincare transformations. As such the Goldstone fields under went shifts under the local symmetry and indeed they  could even be set to zero using this symmetry. 
\par
They also included gauge fields for all the generators of the super Poincare group, but by decorating these with the Goldstone fields they could construct quantities that only  transformed covariantly under the Lorentz transformations. The local translations and supersymmetry transformations of the gauge fields were  absorbed by the Goldstone fields. As such even though they introduced gauge fields for the super Poincare group these did not occur in isolation in the theory but only in combination with the Goldstone fields. As such,  demanding invariance under  the local transformations of the Poincare group could not be used to place strong restriction on  the theory as would usually be the case for a gauge theory. 
\par
In particular  their theory contained a Goldstone fermion $\lambda_\alpha$ and a  gauge field $\psi_{\mu\alpha}$ (gravitino) corresponding to  the supersymmetry  generators  of the super Poincare group. Under a local supersymmetry transformation the gravitino and the Goldstino transformed as $\delta \psi_{\mu\alpha}  =\partial_\mu \epsilon_\alpha  +\ldots $ and $\delta \lambda _\alpha= \epsilon_\alpha +\ldots $ where $+\ldots $ indicates terms which depend on the fields. As such $ \psi_{\mu\alpha}-\partial_\mu \lambda_\alpha +\ldots $ transformed covariantly. This was an illustration of the  Higgs mechanism, the Goldstone fields in the non-linear realisation were eaten by the gauge fields and  in particular the gravitino eat the Goldstone fermion. The result was   to give quantities that transformed only under the Lorentz algebra.
\par
From the above mentioned quantities,  that just transformed under the Lorentz group,  they constructed an action that had three  terms, a Einstein term, a kinetic term for the gravitino and a cosmological term.  However, the local transformations of the theory did not transform these terms into each other and so their relative coefficients were not fixed. Had they constructed the full theory of spontaneously broken supersymmetry,  rather than its low energy effective action, as embodied in the non-linear realisation,  then they would have  discovered supergravity by taking a suitable truncation. Although they could not recover supergravity one can,  with hindsight,  recognise a number of features of the subsequently discovered  supergravity theory  from their  low energy effective action, see reference [B.4] for a further  account of these papers.  
 \par
Ferrara, Freedman and van Niewenhuizen adopted a  pragmatic approach. In their first paper  [15] they proposed  a supergravity action for the graviton field $e_\mu{}^a$ and the gravitino field $\psi_{\mu\alpha}$, which  was given by  
$$
A=\int d^4x\{ {e\over2\kappa^2}R-{1\over 2} \varepsilon^{\mu\nu\rho\kappa} \bar\psi_\mu i\gamma_5\gamma_\nu
D_\rho\big(w(e)\big)\psi_\kappa 
$$
$$
+{\kappa \over 4} \epsilon ^{\lambda\rho \mu\nu} \bar\psi _\lambda \gamma_d \psi_\rho\{ \bar \epsilon\gamma_5\gamma^d D_\nu\big(w(e)\big)\psi_\mu +{i\over 2} \epsilon_\mu {}^{bcd}  D_\nu\big(w(e)\big)\bar \epsilon \gamma_c\psi_b
-D_b\big(w(e)\big)\bar \epsilon \gamma_c\psi_\nu+D_c\big(w(e)\big)\bar \epsilon \gamma_\nu\psi_b \}
$$
$$
 +{\kappa^2\over 32} (det e)^{-1}( \epsilon^{\tau \nu_1\nu_2\mu} \epsilon_{\tau}{}^{\rho_1\rho_2 \lambda} +
 \epsilon^{\tau \nu_1\lambda\mu} \epsilon_{\tau}{}^{\rho_1\rho_2 \nu_2} 
 -\epsilon^{\tau \nu_2\lambda\mu} \epsilon_{\tau}{}^{\rho_1\rho_2 \nu_1} )\bar\psi_{\nu_1}\gamma_{\lambda}\psi_{\nu_2}\bar\psi _{\nu_1}\gamma_{\mu}\psi_{\rho_2} \}
\eqno(2.1.1)$$

% \bar \psi \gamma\psi\to -i \bar \psi \gamma\psi     \gamma_5\to i\gamma_5

and the local supersymmetry transformations 
$$
\delta e^{\ a}_\mu=\kappa\bar\varepsilon\gamma^a\psi_\mu ,\ \  
\delta\psi_\mu=2\kappa^{-1}D_\mu\big(w(e)\big)\epsilon + {\kappa \over 16}\gamma^{ab} (2\bar\psi_\mu\gamma_a\psi_b
+\bar\psi_a\gamma_\mu\psi_b)\epsilon 
\eqno(2.1.2)$$
In these equations 
$$
R=R_{\mu\nu}^{\ \ ab}e_a^{\ \mu}e_b^{\ \nu}, \ \  \ R_{\mu\nu}^{\ \ ab}{\sigma_{ab}\over 4}=[D_\mu \big(w(e)\big),D_\nu \big(w(e)\big)] ,\ \gamma_\mu= e_\mu{}^a\gamma_a   , 
\eqno(2.1.3)$$
 the Lorentz covariant derivative is given by 
$$
 D_\mu\big(w(e)\big)=\partial_\mu+w_{\mu ab}(e) {\gamma^{ab}\over 4} ,
\eqno(2.1.4)$$ 
and the spin connection is the standard one,  
$$
w_{\mu ab}(e)={1\over2}e^\nu_{\ a}(\partial_\mu e_{b\nu}-\partial_\nu
e_{b\mu})-{1\over2}e_b^{\ \nu}(\partial_\mu e_{a\nu}-\partial_\nu
e_{a\mu})-{1\over2}e_a^{\ \rho}e_b^{\ \sigma}(\partial_\rho e_{\sigma
c}-\partial_\sigma e_{\rho c})e_\mu^{\ c}
\eqno(2.1.5)$$
\par
Their paper described in detail how they discovered supergravity. They started from an action which consisted of the usual Einstein term and the known action for a spin three-halves particle coupled to gravity. They then wrote down  putative supersymmetry  transformations. That  for the vierbein was rather clear in that  it should not contain a derivative,  but be linear in the supersymmetry parameter and the gravitino,  have the good Lorentz structure and matching dimensions. In a crucial step the authors realised that  the free spin three-half action had a local symmetry $\delta\psi_{\mu \alpha}=\partial_\mu \eta _\alpha$ and that the spinor $\eta_\alpha$ should be identified in the interacting theory with the supersymmetry parameter $\eta _\alpha= {2\over \kappa} \epsilon_\alpha $, namely . The putative initial  transformation of the gravitino was  then taken to be $\delta \psi_{\mu \alpha}= {2\over \kappa}  D_\mu\big(w(e)\big)\epsilon _\alpha $. 
\par
Working order by order in the gravitational constant, $\kappa$ the  authors  varied the action under the supersymmetry transformations and cancelled the resulting terms using the fact that a term which contained a factor $\partial_\mu \epsilon_\alpha X^\alpha$, where $X^\alpha$ is a function of the fields, could be cancelled by adding $-{\kappa\over 2} \psi_{\mu\alpha }X^\alpha $. The variation of this term lead to new terms in the variation of the action with a higher power of $\kappa$ and these could be cancelled by adding terms to the action as well as to the variation of the fields.  Proceeding order by order in $\kappa$ the authors arrived at the above results. Indeed this  is how the terms cubic and quartic  in the gravitino in the action of   equation (2.1.1) arose and the  terms quadratic in the gravitino  in the variation of the of equation (2.1.2).  Although they did not explicitly state it  they used what is called the Noether method which we will further explain below. 
\par
The action contained terms quadratic in the gravitino and so  it was not easy to verify that the action  was fully  invariant under the supersymmetry transformations. However, in a note added the  authors stated that they had used a computer programme which  showed  that it was indeed invariant under the supersymmetry transformations. 
\par
A bit later Desser and Zumino [16] published a paper which formulated supergravity in first order form, that is, in terms of  the vierbein $e_\mu{}^a$, the gravitino $\psi_{\mu\alpha}$ and the spin connection $\omega^I_{\mu ,a}{}^b$ which were to be treated as independent fields. They took the transformations of the vierbein to be that given in equation (2.1.2) but the transformation of the gravitino to be given by $\delta \psi_{\mu }= 2\kappa^{-1}D_\mu\big(w^I)\big)\varepsilon$ where $\omega^I_{\mu ,a}{}^b$ was the independent spin connection. The variation of this spin connection was chosen just so that the action they proposed was invariant under these supersymmetry transformations. This  action contained an Einstein term built out of the Riemann curvature constructed from the spin connection $\omega^I_{\mu ,a}{}^b$  and a gravitino term which was the Rarita-Schwinger kinetic term with a covariant derivative also using the independent spin connection.  They also made contact with the second order formulation of Ferrara, Freedom and van Nieuwenhuizen. However, their paper did not attempt to show that the transformations of their independent fields  formed a closed algebra even if one used the equations of motion. The first order formulation was to play  little role in the subsequent development of supergravity. 
\par
Ferrara, Freedom and van Nieuwenhuizen then wrote second paper [17]. Inspired by the paper of Desser and Zumino they realised that the supergravity theory  they had discovered could be much more concisely written.  The result still  had as independent fields only the vierbein and the gravitino.  The action and transformations were given by 
$$
A=\int d^4x\left\{{e\over2\kappa^2}\hat R-{1\over2} 
\varepsilon^{\mu\nu\rho\kappa}\bar\psi_\mu i\gamma_5\gamma_\nu
D_\rho\big(w(e,\psi)\big)\psi_\kappa  \right\}
\eqno(2.1.6)$$
and the local supersymmetry transformations 
$$
\delta e^{\ a}_\mu=\kappa\bar\varepsilon\gamma^a\psi_\mu ,\ \  
\delta\psi_\mu=2\kappa^{-1}D_\mu\big(w(e,\psi)\big)\varepsilon
\eqno(2.1.7)$$
In these equations 
$$
\hat R=\hat R_{\mu\nu}^{\ \ ab}e_a^{\ \mu}e_b^{\ \nu}, \ \ \hat R_{\mu\nu}^{\ \ ab}{\gamma_{ab}\over 4}=[D_\mu \big(w(e,\psi)\big)\,D_\nu \big(w(e,\psi)\big) ] ,\ \gamma_\mu= e_\mu{}^a\gamma_a 
\eqno(2.1.8)$$
where the covariant derivative is now given by 
$$
 D_\mu\big(w(e,\psi)\big)=\partial_\mu+\hat w_{\mu ab}(e,\psi)  {\gamma^{ab}\over 4} ,
\eqno(2.1.9)$$ 
with   
$$
\hat w_{\mu ab}(e,\psi)=w_{\mu ab}(e) +{\kappa^2\over4}(\bar\psi_\mu\gamma_a\psi_b
+\bar\psi_a\gamma_\mu\psi_b-\bar\psi_\mu\gamma_b\psi_a)
\eqno(2.1.10)$$
They also showed that these transformations formed a closing algebra {\bf but } only if one used the supergravity equations of motion. 
\par
Since the Noether method played a very important part in the discovery of supergravity and its subsequent development  we will  review it here.    This method arose in the study of gravity, indeed it was found that one could construct Einstein's general relativity in this way. 
The idea is to begin with the linearised  theory,  in this case the action for a spin two particle, the graviton,  in the absence of interactions. This theory is invariant under the rigid Poincare group  transformations  and in particular the transformations $\delta h_{\mu\nu}=\zeta^\lambda \partial_\lambda h_{\mu\nu}$ where the parameter $\zeta^\mu$ is a constant. It is also invariant under the transformation $\delta h_{\mu\nu}= \partial_\mu \xi_\nu+  \partial_\nu \xi_\mu$ which are needed to ensure that the field $h_{\mu\nu}$ just describes  two degrees of freedom in four dimensions. 
\par
To construct the interacting theory one just let the parameter $\zeta^\mu$ of the translations of the Poincare group depend on spacetime. The action is no longer invariant but one can gain an invariant action step by step by identifying $\zeta^\mu={2\over \kappa}  \xi^\mu$ where $\kappa$ is the gravitational coupling constant and adding terms to the transformation of $h_{\mu\nu}$ and the action order by order in $\kappa$. In this way one arrives at Einstein's action and the general coordinate transformations of the graviton, or equivalently the metric. 
\par
The Noether method is not very elegant but it is very powerful and can be applied to find many interacting theories. Given a generic  free theory with a rigid symmetry with parameter $\lambda$, whose indices we suppress, the corresponding Noether current is found by  letting  the parameter $\lambda $   depend on spacetime. The theory is no longer invariant but  its variation leads to a  term which must be of the generic  form $\int dx \partial _\mu\lambda  j^\mu$. The quantity $ j^\mu$ is conserved and is the Noether current for the symmetry. 
\par
To find an interacting theory one lets the parameter $\lambda$ in the free theory depend on spacetime. Then we find that the action varies to give the  
above  term which can be cancelled  by introducing a gauge field $A_\mu$ with the transformation $\delta A_\mu ={1\over g} \partial_\mu \lambda$, where $g$ is a coupling constant,  and adding the term $g\int dx A _\mu  j^\mu$ to the action. One then proceeds order by order in the coupling $g$ adding terms to the action and transformation laws. It often happens that the fields of the original  theory are the required gauge fields, as is the case for gravity and supergravity, and so there is no need to introduce external gauge fields. However, in this case  one must identify the parameter of the original theory with that of the gauge symmetry, as we did above. A review of the Noether method, including its application to Yang-Mills theory and further references  can be found in [B.2]. 
\par
The discovery of supergravity explained so far was incomplete. 

\item{-} A transparent proof that supergravity  was actually invariant under the supergravity transformations was not given.

\item{-}  The supersymmetry transformations only closed if one used the equations of motion of the supergravity theory. 

We will now explain how these  two difficulties were resolved.
\medskip
{\bf 2.2 The invariance} 
\medskip
 The proof of invariance came from an unexpected quarter. In 1976 Chamseddine and West constructed the gauge theory of the super Poincare group and constructed the already discovered supergravity [18]. That this was possible was far from clear as Einstein gravity is not a Yang-Mills theory, that is, it is not of the form of field strength squared. These authors introduced the gauge fields of the super Poincare group $
A_\mu= e_\mu{}^a P_a -{1\over 2} \omega_\mu{}^{ab}J_{ab}+{1\over 2}  \psi_{\mu \alpha} Q^\alpha$ and proceeded just as if it was a Yang-Mills theory. The field strengths were defined by $ [\hat D_\mu,\hat D_\nu ]=-R_{\mu\nu}{}^a
P_a� +{1\over 2}R_{\mu\nu}{}^{ab} J_{ab} +{1\over 2}\Psi_{\mu\nu\alpha} Q^\alpha$ where  $\hat D_\mu =\partial_\mu -A_\mu$ and  are given by 
$$
R_{\mu\nu}{}^a= \partial_\mu e_\nu^a- \partial_\nu e_\mu^a+
\omega_{\mu}{}^a{}_c e_\nu^c -\omega_{\nu}{}^a{}_c e_\mu^c
+{1\over 2}\bar\psi_\mu\gamma^a\psi_\nu,
\quad  R_{\mu\nu}{}^{a b}=\partial _\mu \omega_\nu{}^{ab}
+\omega_\mu{}^{ac}\omega_\nu{}_c{}^{b}- (\mu\leftrightarrow \nu),
$$
$$
\Psi_{\mu\nu}= (\partial _\mu
-{1\over 4}\gamma_{cd}\omega_\mu{}^{cd})\psi_\nu
-(\mu\leftrightarrow \nu)\equiv D_\mu \psi_\nu -(\mu\leftrightarrow \nu)
\eqno(2.2.1)$$
The  fields varied under a gauge transformation with parameter 
$\Lambda = v^a P_a -{1\over 2}  \omega^{ab}J_{ab}+ {1\over 2}\bar \epsilon ^\alpha Q_\alpha$ as  
 $\delta A_\mu= \partial_\mu \Lambda -[A_\mu, \Lambda ]$ and the result is given by 
$$
\delta e_\mu^a= \partial_\mu v^a- \omega^{a}{}_c e_\mu ^c
+\omega_\mu{}^{ac} v _c+{1\over 2}\bar \epsilon\gamma^a\psi_\mu,\quad�
\delta \omega_\mu{}^{ab} =\partial_\mu \omega_{}^{ab}
-(\omega^{ac} \omega_\mu{}_c{}^{b}-\omega^{bc} \omega_\mu{}_c{}^{a}),\quad
$$
$$
\delta \psi_\mu =2(\partial _\mu
-{1\over 4}\gamma_{cd}\omega_\mu{}^{cd})\epsilon 
+{1\over 4}\gamma_{cd}\omega^{cd}\psi_\mu\equiv  D_\mu \epsilon  
+{1\over 4}\gamma_{cd}\omega^{cd}\psi_\mu 
\eqno(2.2.2)$$

The authors then took an unusual step, they demanded that the  field strength for translations vanished 
$$
R_{\mu\nu}{}^a=0
\eqno(2.2.3)$$
This broke the gauge symmetry and in particular the  symmetry parameterised by $v^a$! However,  this condition  allowed one to solve for the spin connection and one finds the result given in equation (2.1.10) and so the theory was now only depended on the veirbein and the gravitino and the transformations were those of supergravity, that is, those in equation (2.1.7).  It had another consequence,  the transformations of  the gauge fields under  gauge transformation for the local translations became general coordinate transformations. 
 \par
Taking the action to be Lorentz invariant and linear in the field strengths one finds a unique  invariant action that is  invariant under the above supersymmetry  transformations subject to the constraint of equation (2.2.3). The result was indeed the supergravity action of equation (2.1.6),  but the good point was that using this method one had shown, using the usual analytic methods,  that it was indeed invariant under supersymmetry. In appendix A3  we give this short derivation. 
\par
It remains to comment on the role of the spin connection which obeyed the constraint of equation (2.2.3).  The variation of the action can be written as 
$$
\delta A= \int d^4x ({\delta A\over \delta e_\mu{}^a} \delta e_\mu{}^a+ {\delta A\over \delta \psi_\mu{}_\alpha}\delta \psi_\mu{}_\alpha+ {\delta A\over \delta \omega_\mu{}^{ab}}\delta \omega_\mu{}^{ab})
\eqno(2.2.4)$$
Since we are in second order formalism, the variation of $ \omega_\mu{}{}^{ab}$ is a function of the vierbein and gravitino and its variation  is just that found by varying the vierbein and graviton upon which it depends.  However, the last term of the variation of the action,  given in equation (2.2.4)  vanishes. as 
$$
 {\delta A\over \delta \omega_\mu{}^{ab}}={e\over 2}R_{\kappa\lambda}^{\quad
c}\left(  e_{c}^{\lambda}\left(  e_{a}^{\mu}e_{b}^{\kappa}-e_{b}^{\mu}%
e_{a}^{\kappa}\right)  +e_{c}^{\mu}e_{a}^{\kappa}e_{b}^{\lambda}\right)=0 
 \eqno(2.2.5)$$
as a consequence of the constraint equation (2.2.3). Thus in effect  the variation of the spin connection does not contribute to the variation of the action in second order form.  Equation (2.2.5) is not an equation of motion but an identity. This last point was discussed  in reference [18] and also in reference [19]. The above method of showing the invariance of supergravity under its supersymmetry transformations could be extended to all supergravity theories and this was indeed how they were shown to be invariant. The above was later called the 1.5 order formalism. 
\par
Restricting to the Poincare group we find a speedy derivation of Einstein gravity as a gauge theory of the Poincar\'e group. We give this derivation in appendix A3. 
\par
 Ali Chamseddine and West  also gauged the OSp(1,4)  group and found the transformations of the vierbein, spin connection and gravitino. The results were reported in thesis of Ali Chamseddine [20]. An action was later found by  MacDowell and Mansouri  [21] which agreed with the then known supergravity theory with a cosmological constant [22].  The othosymplectic approach was further developed in [23]. 
 \par
 The ability to derive theories of gravity from the Yang-Mills procedure has proved very useful in the construction of other theories.  During the period 1977-78  Kaku, Townsend and van Nieuwenhuizen constructed $N=1$ conformal supergravity  [24] using a straightforward application of this method. Following closely  the ideas in reference [18] the authors gauged the super conformal group which has the generators 
 $$
P_\mu ,  J_{\mu\nu},  D, K_\mu, Q_\alpha, S_\alpha
\eqno(2.2.6)$$
As such they introduced corresponding gauge fields $e_\mu{}^a,  \omega_{\mu}{}^{ab}, b_\mu \ f_{\mu a}, \psi_{\mu\alpha}, \phi_{\mu\alpha}$ and constructed their field strengths $R_{\mu\nu}{}^a, R_{\mu\nu}{}^{ab},\ldots , R_{\mu\nu}{}^{\alpha}, \ldots$. They then introduced certain constraints on these curvatures,   as in the Poincar\'e case they  set  $R_{\mu\nu}{}^a=0$, but also $R_{\mu\nu}{}^{\alpha}(\gamma^\mu)_{\alpha}{}^{\beta}=0$ and $\epsilon_{\mu\nu}{}^{\rho\kappa} R_{\rho\kappa}{}^{\alpha}(\gamma^{\mu})_{\alpha}{}^{\beta}=0$, so breaking the gauge symmetry. These enable them to solve for the spin connection $\omega_\mu{}^{ab}$, but also eliminate the gauge field $ \phi_{\mu\alpha}$ in terms of a spin one half field. 
Finally they constructed an invariant action which was quadratic in the remaining curvatures and adjusted the coefficients so that it was invariant subject to the constraints. The gauge field $b_\mu$ was not present as it did not appear in the final action. 
\par
The method of reference [18] was also used to construct higher spin theories by using an infinite dimensional generalisation of the Poincare group.    Although this method   adopts a condition that breaks  the gauge symmetry,  once one took this  unconventional step everything worked very well in many contexts and the desired results emerged in a very quick way. 
 \par
 An account of previous  work on gauge symmetry and gravity was given  in reference [B.3].    Briefly, Sciama showed that the spin connection could be treated as a gauge field for the Lorentz group. There were also quite a few papers discussing gauging the Poincare group,  but they did not mean in a Yang-Mills sense,  rather that the rigid Poincare spacetime translations of the coordinate were taken to be a function of spacetime, namely $x^\mu \to x^\mu + \xi^\mu (x)$. 
  \medskip
{\bf 2.3 The algebra} 
\medskip
There was another problem with the original supergravity theory found by Ferrara,  Freedman and van Nieuwenhuizen. 
The supersymmetry only closed if one used the equations of motion of the supergravity theory. Thus  the closure of  the  supersymmetry transformations   was tied to the supergravity theory and it could not be used to construct another theory such as   the coupling of supergravity to any super matter. Previously  the  symmetry of a theory had a closing algebra which  formed a group that  was independent of that theory.    For example,  the Poincar\'e transformations exist independently   of Maxwell's theory and so could  be used  to construct other Poincar\'e  invariant theories. 
\par
Constructing  the coupling of super matter to the original supergravity was  difficult since this required a new theory whose supersymmetry transformations formed a new algebra. Physicists had to resort to  the  Noether method to construct the coupling of matter to the new supergravity theory.  The coupling of supergravity to the gauge multiplet which contains a spin one and a spin one half particle was found in [25,26,27,28] and  generalisation to Yang-Mills supermulitplet. [28]. The coupling to the chiral (Wess-Zumino) super multiplet which has two spin zero and one spin one half were found in [29,30,31,32,33]. However, the calculations were lengthy and each model had a different supersymmetry algebra.
\par
To find a systematic method of constructing the coupling of supergravity to super matter required  local supersymmetry transformations  that closed in a way that was independent of any particular theory. The Wess-Zumino model exhibited a similar behaviour,  if one did not include its auxilary fields $F$ and $G$. These later fields did not lead to on-shell states but they lead to a closing supersymmetry algebra that was independent of the interactions of the Wess-Zumino model.  It was clear that one required  a good set of auxiliary fields for supergravity. 
\par
 It was quite sometime before those working on supersymmetry realised the simple fact that any supermultiplet had equal numbers of bosonic and fermionic degrees of freedom if the translation operator was a one to one operator. This was despite the fact that the proof of this statement from the supersymmetry algebra was very simple. I forget who,  and at which  conference,  this result was announced at,  but I do remember us all immediately checking it out on our favourite supermultiplets. On-shell the graviton and gravitino had two degrees of freedom and so the numbers matched. However, off shell, and taking account of the gauge symmetries,   the graviton has ${4.5\over 2}-4=6$ degrees of freedom but the gravitino has 4.4-4=12 degrees of freedom. Thus in the off shell theory we required auxiliary fields that had  a net 6 bosonic degrees of freedom. The minimal number of auxiliary fields would be 6 bosonic fields.  Quite soon after the discovery of supergravity   Breitenlohner proposed a set of auxiliary fields that had two fermions, providing 8 degrees of freedom,  as well as  14 bosonic degrees of freedom [34,35]. The construction of the theory involving these fields was complicated and this approach was not pursued further. 
\par
In 1978, two years after the discovery of the original version of supergravity,  an  extension  of this theory  to include 6 bosonic auxiliary fields was found by  Stelle and West [36] and independently by Ferrara and van Nieuwenhuizen [37] as well as by Fradkin and Vasiliev  [38]. These fields were denoted by $M$, $N$ and $b_\mu$  and  this theory did have a closing supersymmetry algebra that was independent of the supergravity theory was found.   It had the action 

$$A=\int d^4x\left\{{e\over2\kappa^2}\hat R-{1\over2}\bar\psi_\mu 
R^\mu-{1\over3}e(M^2+N^2-b_\mu b^\mu)\right\}
\eqno(2.3.1)$$
which was invariant under the supersymmetry  transformations  
$$\eqalignno{
\delta e^{\ a}_\mu&=\kappa\bar\varepsilon\gamma^a\psi_\mu\cr
\delta\psi_\mu&=2\kappa^{-1}D_\mu\big(w(e,\psi)\big)\varepsilon+i\gamma_5\left(b_\mu-{1\over3}\gamma_\mu/\!\!\!b\right)\varepsilon
-{1\over3}\gamma_\mu(M+i\gamma_5N)\varepsilon\cr
\delta M&=-{1\over2}e^{-1}\bar\varepsilon\gamma_\mu
R^\mu-{\kappa\over2}i\bar\varepsilon\gamma_5\psi_\nu
b^\nu-\kappa\bar\varepsilon
\gamma^\nu\psi_\nu
M+{\kappa\over2}\bar\varepsilon(M+i\gamma_5N)\gamma^\mu\psi_\mu\cr
\delta N&=-{e^{-1}\over2}i\bar\varepsilon\gamma_5\gamma_\mu
R^\mu+{\kappa\over2}\bar\varepsilon\psi_\nu b^\nu-\kappa\bar\varepsilon
\gamma^\nu\psi_\nu
N-{\kappa\over2}i\bar\varepsilon\gamma_5(M+i\gamma_5N)
\gamma^\mu\psi_\mu\cr
\delta
b_\mu&={3i\over2}e^{-1}\bar\varepsilon\gamma_5\left(g_{\mu\nu}-{1\over3}\gamma_\mu\gamma_\nu\right)R^\nu+\kappa\bar\varepsilon
\gamma^\nu
b_\nu\psi_\mu-{\kappa\over2}\bar\varepsilon\gamma^\nu\psi_\nu b_\mu\cr
&\quad-{\kappa\over2}i\bar\psi_\mu\gamma_5(M+i\gamma_5N)
\varepsilon-{i\kappa\over4}\varepsilon_\mu^{\
bcd}b_b\bar\varepsilon\gamma_5
\gamma_c\psi_d&(2.3.2)}$$
These supersymmetry  transformations  closed without the use of equations of motion to give the result
$$
[\delta_{\varepsilon_1},\delta_{\varepsilon_2}] =\delta_{\rm
supersymmetry}(-\kappa\xi^\nu\psi_\nu)+\delta_{\rm general\
coordinate}(2\xi_\mu)
$$
$$
+\delta_{\rm Local\
Lorentz}\left(-{2\kappa\over3}\varepsilon_{ab\lambda\rho}
b^\lambda\xi^\rho -{2\kappa\over3}\bar\varepsilon_2\sigma_{ab}
(M+i\gamma_5N)\varepsilon_1+2\xi^dw_d^{\ 
ab}\right)
\eqno(2.3.3)$$
where $\xi_\mu=\bar\varepsilon_2\gamma_\mu\varepsilon_1$. 
\par
One way to see what are the auxiliary fields is to consider the connection between the currents of supersymmetric theories and supergravity fields. The currents in certain  supersymmetric theories  had already been worked out by Ferrara and Zumino [39]. 
The  super multiplet  of currents contained the energy momentum tensor, the super current  and the internal chiral current as well as other objects. Given a theory with a rigid symmetry one can let the parameter of the symmetry become a  function of spacetime and, as explained above, by varying the action  one finds the associated current. Using the Noether method, as also explained above, one finds one can construct a theory with a local symmetry by introducing gauge fields which are in one to one correspondence with the currents. For the case of supergravity, the energy momentum and super current  correspond to the graviton and the gravitino. However the super multiplet of currents also contains the chiral current and two other scalar quantities that have no interpretation in terms of currents. As such we must introduce the additional fields $b_\mu$, $M$ and $N$, that is, the above auxiliary fields. 
\par
 It turns out that  supersymmetric theories can  have different super multiplets of currents which differ in the anomalies  they possess  which in turn leads to a different set of auxiliary fields. Indeed there exists an alternative set of six auxiliary fields called new minimal. It is beyond our remit to discuss these. The correspondence between currents in rigid supersymmetric theories and fields in supergravity theories was realised in [40].
\par
The construction of theories  in general relativity was achieved using the tensor calculus. It relied on the fact that composition of two general coordinate transformations had the same closure no matter on what object it acted. The tensor calculus was constructed out of   tensors which by definition were objects  which transformed in a specific way under general coordinate transformations. It also had  a rule that allowed one to multiply tensors together  to form new tensors and form invariants. 
\par
The discovery of the auxiliary fields allowed the construction of a tensor calculus for supergravity  which made it easy to compute the most general coupling of $D=4,$ $N=1$ supergravity to the most general  super matter. We will now explain how the tensor calculus was constructed [41,42].   There were two matter supermulitplets of interest, the chiral super multiplet $S$,   used in the Wess-Zumino model , and the vector supermulitplet   $V$ used to construct supersymmetric Yang-Mills theories.  The  chiral  multiplet has the field content 
$$
S= ( A, B ; \chi_\alpha ; F , G )
\eqno(2.3.4)$$
where $A$, $B$, $F$ and $ G$   are real  scalar fields and  $\chi_\alpha $ is a  Majorana  spinor. 
While the vector multiplet $V$  has the components 
$$
V=\left(  C ; \zeta_\alpha ; H, K, A_{\mu} ; \lambda_\alpha ; D\right)  
\eqno(2.3.5)$$
where 
$\zeta_\alpha $  and $\lambda_\alpha$ are Majorana spinors and $C,$ $H,$ $K$ and $D$ are scalars. 
\par
The first step in constructing the tensor calculus for supergravity was to  generalise the rigid supersymmetry transformations of these super multiplets  to  be local supersymmetry transformations such that they had the  closure of   equation (2.3.3). The reliable Noether procedure came to the rescue. Once the constant supersymmetry parameter of the rigid supersymmetry transformations was taken to be  function of spacetime the transformations no longer closed as the spacetime derivatives now snagged on these parameters. The first step in the Noether procedure was to cancelled such terms  by adding a term to the transformations involving the gravitino.  Proceeding order by order in the gravitational coupling constant the  resulting local  transformations of the chiral field was found to be given by  
 $$
\delta A =\bar{\varepsilon}\chi;\quad \delta B=i\bar{\varepsilon}\gamma_5\chi
\delta\chi =[F+i\gamma_5G+\hat D \!\!\!\!/ (A+i\gamma_5B)]\varepsilon
$$
$$
\delta F =\bar{\varepsilon}\hat D \!\!\!\!/ \chi -\kappa\bar{\eta}\chi ,\ \ 
\delta G =i\bar{\varepsilon}\gamma_5\hat D \!\!\!\!/ \chi +i\kappa\bar{\eta}\gamma_5\chi
\eqno(2.3.6)$$
where
$$
\hat{D}_aA=\partial_a A-{\kappa\over 2}\bar{\psi}_a\chi,\quad \hat{D}_aB=\partial_aB-{i\kappa\over 2}\bar{\psi}_a\gamma_5\chi ,\ \ 
$$
$$
\hat{D}_a\chi =\left(D_a-{i\kappa\over 2}b_a\gamma_5\right)\chi -{\kappa\over 2}(\hat{D}(A+i\gamma_5B)+F+i\gamma_5G)\psi_a
$$
$$
 \eta =-{1\over 3}(M+i\gamma_5 N+ib  \!\!\!/   ,\gamma_5)\varepsilon
 \eqno(2.3.7)$$
 The corresponding result for the vector multiplet had many similar features but was a bit more complicated. 
\par
 The next step was to find out how to multiply two super multiplets so as to find another super multiplet [41, 42]. In  rigid supersymmetry given two  chiral multiplets $S_1$ and $S_2$ one could form  a chiral super multiplet denoted $S_1\cdot S_2$ but also two  vector multiplet $S_1\times S_2$ and $S_1\wedge S_2$ which are respectively symmetric and anti-symmetric in the exchange of $S_1$ and $S_2$. Given two  vector super multiplets $V_1$ and $V_2$  one can form a third vector multiplet $V_1\cdot V_2$. To generalise these rules to local supersymmetry one must find composite  super multiplets which transform according to the local supersymmetry transformations, such as for the chiral multiplet of equation (2.3.6),  rather than those of rigid supersymmetry.  
 It turns out that the  composition  formulae for local supersymmetry were  the same as in the rigid case except that one has to replace the spacetime derivatives by those that occur in the local case, for example for the chiral fields one takes $\partial_\mu \bullet  \to \hat D_\mu \bullet $ where $\bullet$ is $A$, $B$ or $\chi$. 
 \par
The final step in the construction of the tensor calculus is to construct the supersymmetric invariants for the chiral and vector multiplets.  For rigid supersymmetry these are just given by the integrals over spacetime  of the  $F$ and $D$  fields respectively. Their  generalisation to be invariant under local supersymmetry are easy to find given the local supersymmetry transformations of the fields [41,42]. The invariant for the chiral super multiplet, the $F$ term,  is given by   
$$
[S]_F =\int d^4 x  e \left(  F-\left(  MA+NB\right)  +{1\over 2}\overline{\psi}_{\mu
}\gamma^{\mu}\chi+{1\over 4}\overline{\psi}_{\mu}\gamma^{\mu\nu}\left(
A+i\gamma_{5}B\right)  \psi_{\nu}\right) 
\eqno(2.3.8)$$
While the invariant for the vector super multiplet, the $D$ term,   is given by 
$$
[V]_D  =\int d^4 x  e\{D-{i\kappa\over 2}\overline{\psi}_{\mu}\gamma^{\mu}\gamma
_{5}\lambda+{2\over 3}\left(  MK-NH\right)  -{2\kappa\over 3}A_{\mu}\left(
b^{\mu}+{3\kappa e^{-1}\over 8}\epsilon^{\mu\nu\rho\sigma}\overline{\psi
}_{\nu}\gamma_{\rho}\psi_{\sigma}\right) 
 $$
 $$
 -{\kappa\over 3}\overline{\zeta}\left(  i\gamma_{5}\gamma_{\mu}R^{\mu
}+{3\kappa\over 8}\epsilon^{\mu\nu\rho\sigma}\psi_{\mu}\overline{\psi}_{\nu
}\gamma_{\rho}\psi_{\sigma}\right)  -{2\kappa^{2}\over 3}e^{-1}L_{SG}\}
\eqno(2.3.9)$$
where $L_{SG}$ is the Lagrangian of simple supergravity.   
A complete discussion of the tensor calculus can be found in chapter thirteen  of  the book of reference [B.2]. 
\par
Using the tensor calculus it was straightforward to find the coupling of super matter to supergravity. For example,  the coupling of the chiral multiplet of the Wess-Zumino model to supergravity is given by 
$$
[S\times S]_D-{m\over 2} [S\cdot S]_F +{\lambda \over 3!}[S\cdot S \cdot S]_F
\eqno(2.3.10)$$
While the most general coupling of a scalar field to supergravity is given by 
$$
[f(S) \times S]_D +[g(S)]_F
\eqno(2.3.11)$$
where $f$ and $g$ are arbitrary functions. 
\par
Using the tensor calculus,  the most general coupling of supergravity to the chiral super multiplets was found  in [43,44] and the most general coupling with both the  chiral super multiplets and Yang-Mills super multiplets was found in [45,46,47]. The bosonic part of the action, after the elimination of the   auxiliary fields for supergravity and matter multiplets is given by 

$$
\int d^4 x e\{{1\over 2\kappa^{2}}R-{1\over 4}(Re f_{\alpha \beta} )F_{\mu\nu}^{\alpha}F^{\mu
\nu\alpha}-{1\over \kappa^{2}}{\cal{G}},_{a}{}^{b}D_{\mu}z^{a}D^{\mu}
z_{b} +{i\over 4}(Im f_{\alpha \beta} )\tilde F_{\mu\nu}^{\alpha}F^{\mu
\nu\alpha}
$$
$$
 -{1\over \kappa^{4}}e^{-{\cal{G}}}\left(  3+\left(  {\cal{G}}
^{-1}\right)  _{a}{}^{b}{\cal{G}}_{,}{}^{a}{\cal{G}},_{b}\right)  -
{1\over 8\kappa^{4}}\left\vert g_{\alpha}{\cal{G}}_{,a}\left(  T^{\alpha
}z\right)  ^{a}\right\vert ^{2}\}
\eqno(34)$$
where $F_{\mu\nu}^{\alpha}$ is the Yang-Mills field strength and the function
${\cal{G}}$ is defined by
$$
{\cal{G}}=3\ln\left(  -{\kappa^{2}\over 3}\phi\left( z^{a},z_{a}\right)
\right)  -\ln\left(  {\kappa^{6}\over 4}\left\vert g\left(  z^{a}\right)
\right\vert ^{2}\right)
\eqno(35)$$
The scalar fields $z^a=A^a+iB^a$ are contained in  the chiral super multiplet and they belong to a representation of the gauge group labelled by the index $a$ and $\phi\left( z^{a},z_{a}\right)$ is an arbitrary function. 
The quantity  $D_{\mu}z^{a}$ as covariant derivative with respect to gauge
group $G,$ ${\cal{G}}_{,}^{a}={\partial{\cal{G}}\over \partial z_{a}},$
${\cal{G}},_{a}={\partial{\cal{G}}\over \partial z^{a}},$, 
${\cal{G}},_{a}{}^b={\partial^2{\cal{G}}\over \partial z^{a}\partial z_{b}} $,
$T^{\alpha}$  and the $g_{\alpha}$ are the matrices and gauge couplings associated with the
representation carried by $z^{a}$. The function $f_{\alpha\beta
}\left(  z^{a}\right)  $ is a function of the  chiral fields. 
\par
This theory contains the graviton, the gravitino as well as particles of spins one, one half and zero. It is the
most general supersymmetric theory involving spins two and less which does not have higher spacetime derivatives. It was, and perhaps still is,  the most promising starting point for the construction of realistic models of supersymmetry. The first steps in the construction of such models was taken in reference [48]. 
\par
Akulov, Volkov and Soroka [12] had introduced the appropriate superspace framework for local supersymmetry, but it required a set of   constraints on the torsions and curvatures that lead to the supergravity theory in ordinary spacetime. This was possible as the tangent space group was just the Lorentz group under which the curvatures and torsions belonged to a highly reducible representation. The good constraints were found by Wess and Zumino [49] and this allowed them to indeed recover the supergravity theory. The superspace formalism automatically encoded a supersymmetry algebra that was independent of the construction of any model and as such it  also contained the auxiliary fields. Knowing the auxiliary fields  in ordinary space was no doubt helpful in finding the formulation of supergravity in superspace.  
\par
 Not long after the paper of Wess and Zumino  there appeared a paper by  Warren Siegel claiming that their constraints were so strong that there were no physical degrees of freedom left [50]. Wess and Zumino explained that this paper was wrong but  they had a worried look. To us postdocs it was very amusing that another  young unknown, at least  to us,  post doc had dared to contradict a paper of the gods.  Eventually it was clear that the paper of Warren Siegel was  wrong but he then wrote another  paper [51] solving the constraints on  the  torsions and curvatures. A pedagogical and further elucidation of  this work was given in a paper by Gates and Siegel [52]. To solve the constrains required steps  involving technical leaps of such difficulty that I suspect few have worked through these papers.

%%%%%%%%%%%%%%

\medskip
\centerline {\bf Appendix A More detailed discussion of some papers}
\medskip
In this appendix we will give a more technical account of some of the papers we have discussed. 
\medskip
{\bf Appendix A1.  Volkov and Akulov 1972}
\medskip
As seen from a modern perspective  the authors of this paper  introduced a group element of the super Poincare group $g$ of the form 
$$
g(x)= e^{x^a P_a}e^{\lambda_\alpha (x) Q^\alpha}\tilde h
\eqno(A1.1) $$
where $P_a$ and $Q_\alpha$ are the generators of spacetime translations and supersymmetry transformations and $\tilde h$ is an element of the  Lorentz . We note that the parameter of the supersymmetry generator was taken to depend on spacetime and so it was a  field rather than a parameter. The theory was to be invariant under the transformations 
$$
g(x)\to g_0 g(x) ,\ {\rm and } \ g(x)\to g(x) h(x)
\eqno(A1.2)$$
where $g_0 $ was an arbitrary super Poincare group element, but one that did not depend on $x^\mu$ and $h(x)$ was a Lorentz group element which did depend in an arbitrary way on $x^\mu$. Using this latter transformation one could remove the $\tilde h(x)$ part of the group element of equation (A1.1).  
\par
The Cartan forms are given by 
$$
g(x)^{-1}\partial_\mu g(x) = f_\mu {}^a P_a +  k_{\mu \alpha }Q^\alpha
\eqno(A1.3)$$
They only transform under the local Lorentz transformations and the veirbein $f_\mu {}^a = \delta _\mu {}^a -\bar \lambda\gamma^a\partial_\mu \lambda $  and $k_{\mu\alpha} = \partial_\mu \lambda_\alpha$ transforms as their indices suggest. As such an invariant action is given by 
$$
A=\int d^4 x \det f_\mu {}^a
\eqno(A1.4)$$
This contains the usual kinetic term for the spin half particles as well as interactions. 
\par
Volkov was motivated  by the work of Werner Heisenberg in the 1950's which took the attitude that all massless particles were Goldstone particles and in particular the neutrino was a Goldstone particle. For an interesting account  of this history which includes an account in Volkov's own words see reference [B.5]. 
\medskip
{\bf Appendix A2.  Soroka and Volkov 1973}
\medskip
These papers  constructed a non-linear realisation which corresponded to the breaking of the  local super Poincare group to the Lorentz group.  This was achieved by introducing gauge fields of the super Poincare  group in addition to the Goldstone fields. They began with a group element of the super Poincare group of the form 
$$
g(x)= e^{\zeta(x)^a P_a}e^{\lambda^\alpha(x) Q_\alpha} \tilde h(x) 
\eqno(A2.1) $$
where $ \tilde h(x) $ is a Lorentz group element. The group element $g$ transforms as $g(x) \to l (x) g(x) h(x)$ where $l(x) $ and $h(x)$ are local transformations which belongs to the super Poincare group and the Lorentz group respectively. The  fields $\zeta^a$ and $\lambda^\alpha$ are the Goldstone fields corresponding to  translations and supersymmetry. As they now have  local super Poincare transformations they could  be used to gauge away these Goldstone fields. 
\par 
They also introduced gauge fields $A_\mu= e_\mu {}^aP_a + \psi_{\mu \alpha}Q^\alpha + \omega_\mu {}^{ab} J_{ab}$ which transform as 
$$
A_\mu ^\prime = l (x)A_\mu l(x) ^{-1} +l (x)\partial_\mu l(x)^{-1}
\eqno(A2.2)$$
where $l$ is the same group element as in  equation (A2.1).  They then constructed the quantity 
$$
\tilde A_\mu\equiv  g^{-1}A_\mu g+ g^{-1}\partial_\mu  g
\eqno(A2.3)$$
where $g(x)$ is the group element of equation (A2.1) and so it contains the Goldstone fields. It is easy to see that $\tilde A_\mu$ only transforms under the Lorentz group 
$$
\tilde A_\mu ^\prime = h (x) \tilde A_\mu  h(x) ^{-1} +h (x)\partial_\mu h(x)^{-1}
\eqno(A2.4)$$
 Corresponding to this restricted transformation one can construct an action consisting of three terms: Einstein gravity, a kinetic energy term for the graviinto and a cosmological term, but these  do not transform into each other under the above transformations. 
 \par
 In this construction the gauge fields and the Goldstone fields are bound together in such a way that the composite objects just transforms under the Lorentz group. If one were to set the Goldstone fields to zero the gauge fields would, now by themselves, transform under the full symmetry and the construction of an invariant action would be completely different. As such  the gauge fields do not occur in this theory as they do in a Yang-Mills theory. Nonetheless, in these calculations one can, with hindsight,  recognise the vierbein, the gravitino and the spin connection of supergravity but the presence of the Goldstone fields means that the symmetry between them is not realised in the same  way as  that which occurs in the supergravity theory. 
\medskip
{\bf Appendix A3.3  Chamseddine and West  1976}
\medskip
A proof of the  invariance of the supergravity action of Ferrara, Freedman and Van Nieuwenhuizen was given in reference [18]. 
Since it is rather short and key to development of supergravity we give it here. We begin with   the action of equation (2.1.6), when written in terms of the field strength for the gauge fields,  and vary it under the supersymmetry transformation of equation (2.2.2), which are the same as those of  equation (2.1.7),  subject to the  condition of equation (2.2.3).  As explained in section (2.2) we do not need to vary the spin connection as the result is proportional to $R^a_{\mu\nu}=0$ which is the correct equation giving the spin connection in terms of the vierbein. Thus variation of the Einstein term  is given by
the Einstein field equation times the variation of the veirbein  
$$
\delta \int {e\over 2 \kappa^2} \big(e_a{^\mu} e_b{^\nu}  R_{\mu\nu}{^{ab}}\big)\,d^4
x  = \int d^4 x\!\left\{{1\over \kappa} \{\bar
{\varepsilon}\gamma^\mu\psi_a
\}\bigg\{-\! R_\mu{^a}+{1\over 2}e_\mu{^a} R\bigg\}\!\right\}
\eqno(A3.1)$$
The variation of the Rarita-Schwinger part of the action gives the following three terms
$$
\delta \int
\left(-{i\over 2}\bar{\psi}_\mu\gamma_5 e_\nu{}^a\gamma_a D_\rho
\psi_\kappa\varepsilon^{\mu\nu\rho\kappa}\right)\,d^4 x = \int d^4
x\bigg\{-{i\over \kappa}\bar{\varepsilon}
\overleftarrow{D}_\mu\gamma_5\gamma_\nu D_\rho\psi_\kappa
\varepsilon^{\mu\nu\rho\kappa} 
$$
$$
-{i\over \kappa}\bar{\psi}_\mu\gamma_5\gamma_\nu\overrightarrow{D}_\rho
D_\kappa \varepsilon
\varepsilon^{\mu\nu\rho\kappa}-{\kappa\over 2}
i\bar{\varepsilon}\gamma^a \psi_\nu\bar{\psi}_\mu \gamma_5
\gamma_a D_\rho \psi_\kappa
\varepsilon^{\mu\nu\rho\kappa}\bigg\}
\eqno(A3.2)$$
Flipping the spinors using their Majorana property we find that the second term of the above equation takes the form
$$
-{i\over 8 \kappa}\bar{\psi}_\mu \gamma_5
\gamma_\nu R_{\rho\kappa}{}^{cd}
\sigma_{cd}\varepsilon\varepsilon^{\mu\nu\rho\kappa} =
-{i\over 8 \kappa}
\bar{\varepsilon}\sigma_{cd}\gamma_\nu\gamma_5 \psi_\mu
R_{\rho\kappa}{^{cd}}\varepsilon^{\mu\nu\rho\kappa}
\eqno(A3.3)$$
Integrating the first term of Eq. (A3.2) by parts and neglecting surface terms we find that it is given by 
$$
 {i\over \kappa}\bar{\varepsilon}\gamma_5 [D_\mu,
\gamma_\nu]D_\rho\psi_{\kappa}\varepsilon^{\mu\nu\rho\kappa}
+{i\over \kappa}\bar{\varepsilon}\gamma_5\gamma_\nu D_\mu
D_\rho\psi_\kappa\varepsilon^{\mu\nu\rho\kappa}
\eqno(A3.4)$$
The second of these terms is given by 
$$
{i\over 8 \kappa}\bar{\varepsilon}\gamma_5\gamma_\nu
R_{\rho\kappa}{^{cd}} \sigma_{cd}\psi_\mu
\varepsilon^{\mu\nu\rho\kappa}
\eqno(A3.5)$$

The term  given in equation (A3.5) and that in equation (A3.3) add together to give the result
$$
+{i\over 2\cdot 4 \kappa}\bar{\varepsilon}\gamma_5(\gamma_\nu
\sigma_{cd} + \sigma_{cd}\gamma_\nu)\psi_\mu
R_{\rho\kappa}{^{cd}}\varepsilon^{\mu\nu\rho\kappa}
={1\over 4 \kappa}
\bar{\varepsilon}\gamma_f\psi_\mu\varepsilon_{f\nu c
  d}\varepsilon^{\mu\nu\rho\kappa}R_{\rho\kappa}{}^{cd}
$$
$$
=-{1\over 2\kappa}\bar{\varepsilon}\gamma^a\psi_\mu\big\{e_a{^\mu}
R - 2 R_a{^\mu}\big\} e
\eqno(A3.6)$$
which  exactly cancels the variation of the Einstein~action given in Eq. (A3.1).

We are  left with the first term of equation (A3.4) and the last term of equation  (A3.2), Performing a Fierz transformation on the latter term it becomes
$$
-{\kappa\over 2\cdot 4}i\bar{\varepsilon}\gamma^a \gamma_R
\gamma_a \gamma_5 D_\rho
\psi_\kappa\varepsilon^{\mu\nu\rho\kappa}\bar{\psi}_\mu\gamma_R\psi_\nu = +{\kappa\over 4}
i\bar{\varepsilon}\gamma_c\gamma_5D_\rho\psi_\kappa\varepsilon^{\mu\nu\rho\kappa}\bar{\psi}_\mu\gamma^c\psi_\nu
\eqno(A3.7)$$
The first term in  equation (A3.4) is most easily  evaluated by going to inertial coordinates,
that is, we set $\partial_\mu e_\nu{^a} = 0$; it becomes
$$
{i\over 4 \kappa} \bar{\varepsilon}\gamma_5
[\sigma^{cd},\gamma_\nu]
w_{\mu cd}D_\rho\psi_\kappa\varepsilon^{\mu\nu\rho\kappa}={i\over \kappa}\bar{\varepsilon}\gamma_5\gamma^c D_\rho\psi_\kappa w_{\mu c\nu} \varepsilon^{\mu\nu\rho\kappa}
$$
$$
={\kappa\over 4}i\bar{\varepsilon}\gamma_5 \gamma^c D_{\rho}
  \psi_{\kappa} \bar{\psi}_\mu \gamma_c \psi_\nu
  \varepsilon^{\mu\nu\rho\kappa}
  \eqno(A3.8)$$
  This term cancels with that of equation  (A3.7). This completes the proof of invariance.
\par
Clearly gauging the Poincare group to find gravity is a sub case of the discussion of reference [18]. Although this was not spelt out in that reference we give it in detail here in order to demonstrate just how quickly one can derive Einstein's gravity from the gauge viewpoint.  
The field strengths of the Poincare group are 

$$
R_{\mu\nu}{}^a= \partial_\mu e_\nu^a- \partial_\nu e_\mu^a+
\omega_{\mu}{}^a{}_c e_\nu^c -\omega_{\nu}{}^a{}_c e_\mu^c \equiv D_\mu e_\nu{}^a-D_\nu e_\mu{}^a
\eqno(A3.9)$$
and 
$$
R_{\mu\nu}{}^{a b}=\partial _\mu \omega_\nu{}^{ab}
+\omega_\mu{}^{ac}\omega_\nu{}_c{}^{b}- (\mu\leftrightarrow \nu)
\eqno(A3.10)$$
\par
The gauge transformations of the fields are given by 
$$
\delta e_\mu^a= \partial_\mu v^a- \omega^{a}{}_c e_\mu ^c
+\omega_\mu{}^{ac} v _c \equiv D_\mu v^a,\quad
\delta \omega_\mu{}^{ab} =\partial_\mu \omega_{}^{ab}
-(\omega^{ac} \omega_\mu{}_c{}^{b}-\omega^{bc} \omega_\mu{}_c{}^{a}) ,\quad
\eqno(A3.11)$$
\par
We adopt the constraint 
$$
R_{\mu\nu}{}^a=0
\eqno(A3.12)$$
which expresses the spin connection in terms of the vierbein, so moving to second order formalism. 
\par
The most general action one can write out of the remaining field strength and the vierbein is the Einstein action which we rewrite 
$$
A= {1\over 2 \kappa^2} \int d^4 x e e_a{}^\mu e_b{}^\nu R_{\mu\nu}{}^{ab}= -{1\over 4 \kappa^2 }\int d^4 x \epsilon ^{\mu\nu\rho\kappa} \epsilon _{abcd} e_\mu {}^a e_\nu{}^b R_{\rho\kappa }{}^{cd}
\eqno(A3.13)$$
Varying the action under the gauge transformations of equation (A3.11) we get 
$$
\delta A= -{1\over 4 \kappa^2 }\int d^4 x \epsilon ^{\mu\nu\rho\kappa} \epsilon_{abcd}\{
2D_\mu v^a e_\nu{}^b R_{\rho\kappa }{}^{cd}+ e_\mu {}^a e_\nu{}^b \delta R_{\rho\kappa }{}^{cd} \}
\eqno(A3.14)$$
Integrating by parts we have 
$$
-{1\over 4 \kappa^2 }\int d^4 x \epsilon ^{\mu\nu\rho\kappa} \epsilon_{abcd}\{-2 v^a D_\mu e_\nu{}^bR_{\rho\kappa }{}^{cd}
-2 v^a e_\nu{}^b D_\mu R_{\rho\kappa }{}^{cd} + e_\mu {}^a e_\nu{}^b \delta R_{\rho\kappa }{}^{cd} \}
 \eqno(A3.15)$$
 The second term vanishes by the Bianchi identity. Both the remaining terms are given by  
 $$
 \delta A= {1\over 4 \kappa^2 }\int d^4 x \epsilon ^{\mu\nu\rho\kappa} \epsilon_{abcd}\{-v^a R_{\mu\nu}{}^bR_{\rho\kappa }{}^{cd}
 +2 R_{\rho\mu }{}^a e_\nu{}^b \delta \omega_\kappa {}^{cd} \}
  \eqno(A3.16)$$ 
 These  vanish due to the constraint of equation (A3.12) and so we find, subject to the constraint of equation (A3.12), an invariant theory. We observe that  the variation of the spin connection in the action vanishes due to this contstraint no matter what is its variation. Just why the gauge approach to find theories of gravity  is so very efficient is not really clear.

 %%%%%%%%%
 
\medskip
{\bf{ Acknowledgments}}
I wish to thank Dio Anninos, Ali Chamseddine, Misha Shifman and Dmitri Sorokin  for discussions and help while preparing this manuscript. 
\medskip
{\bf References}
\medskip
\item{[1]} Yu. A. Golfand and E. P. Likhtman, {\it Extension of the algebra of Poincar«e group generators and
violation of P invariance}, JETP Lett. 13, 323 (1971) [Pisma Zh. Eksp. Teor. Fiz. 13, 452 (1971)].
\item{[2]} S. Coleman and J. Mandula, {\it All Possible Symmetries of the S Matrix},  Phys. Rev. 159 (1967): 1251. 
\item{[3]}  D. V. Volkov and V. P. Akulov, {\it Possible universal neutrino interaction,}  JETP Lett. 16, 438 (1972) which also appeared as {\it Is the neutrino a Goldstone particle?},  Phys. Lett. B 46, 109 (1973).  
\item{[4]} D. V. Volkov and V. P. Akulov,  {\it  Goldstone fields with spin one half}, Theor. Math. Phys. 18, 28 (1974)
28 [Teor. Mat. Fiz. 18, 39 (1974)].
\item{[5]} A. Salam and J. A. Strathdee, {\it Super-gauge transformations}, Nucl. Phys. B 76, 477 (1974); {\it On Superfields and Fermi-Bose Symmetry}, Phys. Rev. {\bf D11} (1975) 1521.
\item {[6]} ÊÊP. Ramond, {\it Dual theory for free fermions}, Phys. Rev. {\bf ÊD3} (1971) 2415.
\item {[7]} A. Neveu and J.H. Schwarz, {\it Factorizable dual model of pions}, Nucl. Phys. {\bf B31} (1971) 86;  A. Neveu and J.H. Schwarz, {\it Quark Model of Dual Pions}, Phys. Rev.{\bf  D4} (1971) 1109. 
\item {[8]} J. L. Gervais and B. Sakita, {\it Field Theory Interpretation Of Supergauges In Dual Models}, Nucl. Phys. {\bf B34} (1971)
632.
\item{[9]}  B. Zumino, {\it Relativistic Strings and Supergauges} in Renormalization and Invariance in Quantum Field theory, edited by E. Caianiello, page 367, Plenum Press 1974.
\item{[10]} J. Wess and B. Zumino, {\it Supergauge transformations in four dimensions}, Nucl. Phys.
{\bf B70} (1974) 139;  {\it A Lagrangian Model Invariant Under Supergauge Transformations},
Phys. Lett. {\bf 49B} (1974) 52.
\item{[11]} R. L. Arnowitt, P. Nath and B. Zumino, {\it Superfield densities and action principle in
curved superspace}, Phys. Lett. B 56 (1975) 81Ð84
\item{[12]} D. V. Volkov, V. P. Akulov and V.A. Soroka, {\it Gauge fields on superspaces with different holonomy groups}, Theor.
Math. Phys. 20, 829 (1974) [Teor. Mat. Fiz. 20, 291(1974)]. 
\item{[13]} D. V. Volkov and V. A. Soroka, {\it Higgs effect for Goldstone particles with spin 1/2}, JETP Lett.
18, 312 (1973) [Pisma Zh. Eksp. Teor. Fiz. 18, 529 (1973)]. 
\item{[14]}  D. V. Volkov and V. A. Soroka, {\it Gauge fields for symmetry group with spinor parameters}, Theor.
Math. Phys. 20, 829 (1974) [Teor. Mat. Fiz. 20, 291(1974)].
\item{[15]} D. Z. Freedman, P. van Nieuwenhuizen and S. Ferrara, {\it Progress toward a theory of
supergravity}, Phys. Rev. D13 (1976) 3214Ð3218
\item{[16]} S. Deser and B. Zumino, {\it Consistent Supergravity}, Phys. Lett. 62B (1976) 335
\item{[17]} D. Z. Freedman, P. van Nieuwenhuizen and S. Ferrara,  {\it Properties of supergravity theory}, Phys. Rev. {\bf D14}, 912 (1976). 
\item{[18]} A. Chamseddine and P. West, {\it Supergravity as a gauge theory of supersymmetry}, {\it Nucl. Phys.} {\bf
B129}, 39 (1977). This paper was received on the 28 September 1976 but was initially rejected for publication. The  revised version,  received  on  10 June 1977, was essentially the same except for two paragraphs explaining in more detail its relation to references [15] and [17]. It was also appeared  as the  Imperial preprint ICTP/75/22, September 1976. 
\item {[19]} P. Townsend and P. van Nieuwenhuizen, {\it Geometrical interpretation of extended supergravity},  {\it Phys.
Lett.} {\bf B67}, 439 (1977). 
\item{[20]} A. H. Chamseddine, {\it Supersymmetry and higher spin fields},  Ph.D. Thesis, defended September 1976, Imperial College, London University. Chapter 6 of the thesis is stated as collaborative work with P. West. 
The thesis can be found at 
\par
https://drive.google.com/file/d/0B8tITtoqQkxfeHQ3cFJ0WUxIcHM/view?usp=sharing 
  It's contents were also discussed with a number of physicists  who passed through Imperial during 1976 . 
\item{[21]} S. MacDowell and F. Mansouri, {\it Unified geometric theory of gravity and supergravity}, Phys. Rev. Lett. {\bf 38 }(1977) 739. This paper was received on the 9 February 1977. 
\item{[22]} Paul K. Townsend, {\it Cosmological constant in supergravity}, Phys. Rev. {\bf D15} (1977) 2802. 
\item {[23]} Ali Chamseddine, {\it Massive Supergravity from Spontaneously Breaking Orthosymplectic Gauge Symmetry},   Annals Phys. 113 (1978) 219; {\it Massive Supergravity from Nonlinear Realization of Orthosymplectic Gauge Symmetry and Coupling to (Spin 1/2, Spin 1) Multiplet},  Nucl. Phys.B 131 (1977) 494. 
\item{[24]} P. Van Nieuwenhuizen, M. Kaku, and P. Townsend, {\it Properties of conformal supergravity}, Phys. Rev. {\bf D17}, (1978) 1501; {\it Superconformal Unified Field Theory}, Phys. Rev. Lett. {\bf 39 }(1977) 1109. 
\item{[25]} S. Ferrara, J. Scherk and P. van Nieuwenhuizen, {\it Locally supersymmetric
Maxwell-Einstein theory}, Phys. Rev. Lett. 37 (1976) 1035
\item{[26]} D. Z. Freedman, {\it Supergravity with axial gauge invariance}, Phys. Rev. D15 (1977)
1173.
\item{[27]} B. de Wit, {\it On a locally supersymmetric extension of quantum electrodynamics}, Phys.
Lett. B 66 (1977) 77
\item{[28]} D. Z. Freedman and J. H. Schwarz, {\it Unification of supergravity and Yang-Mills theory},
Phys. Rev. D15 (1977) 1007.
\item{[29]} S. Ferrara, F. Gliozzi, J. Scherk and P. Van Nieuwenhuizen, {\it Matter couplings in
supergravity theory}, Nucl. Phys. B117 (1976) 333
\item{[30]} S. Ferrara, D. Z. Freedman, P. van Nieuwenhuizen, P. Breitenlohner, F. Gliozzi and
J. Scherk, {\it Scalar multiplet coupled to supergravity}, Phys. Rev. D15 (1977) 1013
\item{[31]} E. Cremmer and J. Scherk, {\it Modified interaction of the scalar multiplet coupled to
supergravity}, Phys. Lett. B 69 (1977) 97Ð100
\item{[32]} A. K. Das, M. Fischler and M. Rocek, {\it Massive, selfinteracting scalar multiplet
coupled to supergravity}, Phys. Lett. B 69 (1977) 186Ð188
\item{[33]} A. Das, M. Fischler and M. Roÿcek, {\it Super-Higgs effect in a new class of scalar models
and a model of super QED}, Phys. Rev. D16 (1977) 3427Ð3436.
\item{[34]} P. Breitenlohner, {\it A geometric interpretation of local supersymmetry}, Phys. Lett. B67
(1977) 49
\item{[35]} P. Breitenlohner, {\it Some invariant Lagrangians for local supersymmetry}, Nucl. Phys.
B124 (1977) 500
\item{[36]} K. Stelle and P. West, {\it Minimal auxiliary fields for supergravity}, {\it Phys. Lett.} {\bf B74}, 569. 
330 (1978).   
\item{[37]} S. Ferrara and P. van Nieuwenhuizen, {\it The auxiliary fields of supergravity}, Phys.
Lett.{\bf B74}, 333 (1978).
\item{[38]} E. S. Fradkin and M. A. Vasiliev, {\it S matrix for theories that admit closure of the
algebra with the aid of auxiliary fields: the auxiliary fields in supergravity}, Nuovo
Cim. Lett. 22 (1978) 651.
\item{[39]} S. Ferrara and B. Zumino,  {\it  Transformation properties of the supercurrent}, Nuclear Physics {\bf B87}(1975) 207
\item{[40]} V. Ogievetski and E Sokatchev, {\it n Vector Superfield Generated by Supercurrent} , Nucl. Phys. {\bf B124} (1977) 309. 
 \item{[41]}  K.S. Stelle and P. West, {\it Tensor calculus for the vector multiplet coupled to supergravity},  {\it Phys. Lett.} {\bf 77B},
376 (1978);  {\it Relation between vector and scalar multiplets and gauge invariance in supergravity},  Nucl. Phys. {\bf B145} (1978) 175.
\item{[42]} S. Ferrara and P. van Nieuwenhuizen, {\it Tensor calculus for supergravity}, Phys. Lett. {\bf 76B} (1978) 404;  {\it Structure of supergravity}, Phys. Lett. B78 (1978) 573
\item{[43]}  E. Cremmer, B. Julia, J. Scherk, P. van Nieuwenhuizen, S. Ferrara and L. Girardello,
{\it Super-higgs effect in supergravity with general scalar interactions}, Phys. Lett. B79
(1978) 231
\item{[44]}  E. Cremmer, B. Julia, J. Scherk, S. Ferrara, L. Girardello and P. van Nieuwenhuizen,
{\it Spontaneous symmetry breaking and Higgs effect in supergravity without cosmological
constant}, Nucl. Phys. B147 (1979) 105
\item{[45]}  E. Cremmer, S. Ferrara, L. Girardello and A. Van Proeyen,
{\it Coupling supersymmetric Yang-Mills gauge theories to supergravity}, Phys. Lett. {\bf 116B} (1982) 231.
\item{[46]}  E. Cremmer, S. Ferrara, L. Girardello and A. Van Proeyen, {\it YangÐMills theories with
local supersymmetry: Lagrangian, transformation laws and superhiggs effect}, Nucl.
Phys. B212 (1983) 413Ð442
\item{[47]}  P. Nath, R. Arnowitt and A. H. Chamseddine, {\it Applied N=1
Supergravity, }ICTP\ series in Theoretical Physics, Volume 1, (1982), World
Scientific, Singapore.   
\item{[48]} A. H. Chamseddine, R. L. Arnowitt and P. Nath, {\it Locally supersymmetric Grand
Unification}, Phys. Rev. Lett. 49 (1982) 970 
\item{[49]}  J. Wess and B. Zumino, {\it Superspace formulation of supergravity}, Phys Lett {\bf 66B} (1977) 361
\item{[50]} W. Siegel,  {\it A Comment on the Wess-Zumino Formulation of Supergravity} 1977, Harvard University preprint.  \item{[51]} W. Siegel, {\it Solution to Constraints in {Wess-Zumino} Supergravity Formalism},     Nucl.Phys. {\bf B142} (1978) 301-305; 
 {\it Supergravity Superfields Without a Supermetric}, Harvard preprint HUTP-771 A068, {\it Nucl. Phys.} {\bf B142}, 301 (1978); 
\item{[52]} S.James Gates, Jr and W. Siegel, {\it Understanding Constraints in Superspace Formulations of Supergravity},    Nucl.Phys. {\bf B163} (1980) 519.

%%%%%%%%%%

\medskip
{\bf Additional pedagogical references }
\medskip
\item{[B.1]} G. Kane and M. Shifman, {The supersymmetric world, the beginnings of the theory}, second edition, World Scientific, 2000. 
\item{[B.2]} A discussion of the Noether method applied to Yang-Mills theory can be found in chapter 7 in P. West, {\it Introduction to supersymmetry and supergravity},  World Scientific, 1990 and to gravity and supergravity in chapter 13 of {\it Introduction to Strings and Branes}, Cambridge 2012. 
\item{[B.3]} A. Chamseddine and P. West, {\it The role of the 1.5 order formalism and the gauging of  spacetime groups   in  the development of   gravity and supergravity theories}, Mod.Phys.Lett. {\bf A 37} (2022) 08, 2230005, arXiv:2201.06874. 
\item{[B.4]} S. Kuzenko,  {\it Local supersymmetry: Variations on a theme by Volkov and Soroka},     Proc.Roy.Soc.Lond.A 479 (2023) 2271, 20230022, [hep-th] arXiv:2110.12835
\item{[B.5]} M. Shifman, {\it ÒFrom Heisenberg to Supersymmetry} , Forsch. Physics. 50 (2002), 552. 

\end